\documentclass[aps,pre,twocolumn,showpacs,amsmath,amssymb,superscriptaddress]{revtex4-1}
\usepackage{amsmath}
\usepackage{graphicx}
\usepackage{verbatim}
\usepackage{color}
\usepackage{subfigure}
\usepackage{hyperref}
\usepackage{soul}
\usepackage{amssymb}
\usepackage{cancel}
\usepackage{float}
\begin{document}
\title{Magnetic particles confined in a modulated channel: structural transitions tunable by 
  tilting a magnetic field}

\author{J. E. \surname{Galv\'an-Moya}} \email[Email: ]{JesusEduardo.GalvanMoya@uantwerpen.be}
\affiliation{Department of Physics, University of Antwerp, Groenenborgerlaan 171, B-2020, Antwerp, Belgium}

\author{D. \surname{Lucena}}
\affiliation{Department of Physics, University of Antwerp, Groenenborgerlaan 171, B-2020, Antwerp, Belgium}
\affiliation{Departamento de F\'isica, Universidade Federal do Cear\'a, Caixa Postal 6030, Campus do Pici, 60455-760 Fortaleza, Cear\'a, Brazil}

\author{W. P. \surname{Ferreira}}
\affiliation{Departamento de F\'isica, Universidade Federal do Cear\'a, Caixa Postal 6030, Campus do Pici, 60455-760 Fortaleza, Cear\'a, Brazil}

\author{F. M. \surname{Peeters}} \email[Email: ]{Francois.Peeters@uantwerpen.be}
\affiliation{Department of Physics, University of Antwerp, Groenenborgerlaan 171, B-2020, Antwerp, Belgium}
\affiliation{Departamento de F\'isica, Universidade Federal do Cear\'a, Caixa Postal 6030, Campus do Pici, 60455-760 Fortaleza, Cear\'a, Brazil}

\date{\today}

\begin{abstract}
The ground state of colloidal magnetic particles in a modulated channel are investigated as
 function of the tilt angle of an applied magnetic field.  The particles are confined by a
 parabolic potential in the transversal direction while in the axial direction a periodic
 substrate potential is present.  By using Monte Carlo (MC) simulations, we construct a phase
 diagram for the different crystal structures as a function of the magnetic field orientation,
 strength of the modulated potential and the commensurability factor of the system.
 Interestingly, we found first and second order phase transitions between different crystal
 structures, which can be manipulated by the orientation of the external magnetic field.
 A re-entrant behavior is found between two- and four-chain configurations, with continuous
 second order transitions.  Novel configurations are found consisting of frozen in solitons.
 By changing the orientation and/or strength of the magnetic field and/or the strength and the
 spatial frequency of the periodic substrate potential, the system transits through different
 phases.
\end{abstract}
\maketitle

\section{Introduction}
Competitive interactions between particles are responsible for the formation of a large variety of
 complex structures in nature\cite{072_shevchenko, 070_min, 071_choueiri}.  However, to take
 advantage of this aspect, the major challenge is to find a way to manipulate this formation
 process, in order to obtain the desired configuration through a correct choice of a set of
 parameters.  Although this idea seems to be promising, there are many other obstacles to
 overcome.

Formation of Wigner crystals has been widely studied for purely repulsive interaction between
 particles in quasi-one-dimensional (Q1D)\cite{003_piacente, 041_galvan, 054_galvan}, 
 two-dimensional (2D)\cite{004_schweigert,009_partoens} and
 three-dimensional (3D)\cite{008_hasse,005_cornelissens} systems. These studies showed a very rich
 set of ground state configurations depending on the system properties, and a diverse kind of
 transitions was found between them.  However, the formation of complex structures increases
 dramatically when particles interact via competing
 interaction\cite{089_reichhardt, 073_liu, 063_drocco, 075_zhao}.  One of the most simple models
 is based on the dipole-dipole interaction\cite{055_froltsov, 084_lowen}, which can be realized by
 an external field acting on a system of magnetic particles.  The multiple crystalline structures
 which can be found\cite{085_reinmuller, 086_deutschlander}, and the effects of external fields
 applied on these particles which have been summarized in a recent mini-review\cite{087_lowen},
 indicate the promising future of these kind of systems.
 
Recent works show the effect of a fixed and oscillating external magnetic field on the crystal
 structures\cite{068_danilov, 066_jager}, evidencing that the configuration of the system can be
 controlled by the orientation of the magnetic field.  Numerical works on these systems
 demonstrated the possibility of cluster formation\cite{065_lucena, 058_knap} as well as few-body
 bound states\cite{056_volosniev, 062_wunsch} and local deformations in elongated dipolar
 gases\cite{069_ruhman}.  Additionally, different transitions between Fermi liquid, solitons and
 Wigner crystals have been predicted\cite{059_spivak, 060_bauer} as well as some tunable
 assemblies\cite{078_yang}.

The effect of complex potential landscapes on the structure of interacting colloidal particles
 was investigated in 1D\cite{077_hanes, 006_kwinten, 018_carvalho} and recently also in
 2D\cite{088_neuhaus}.  A periodic potential has been proposed to control the arrangement of the
 particles\cite{061_carr, 063_drocco, 088_neuhaus}, which is triggered by the
 commensurability\cite{057_herrera, 074_reichhardt} between the periodicity of the potential and
 the average distance, i.e. the density, of the particles.  Experimentally, it was found that a
 colloidal suspension, where the interaction between particles is negligible, exhibits a
 non-trivial dependence on the strength of the modulated potential\cite{076_dalle}. This finding
 allows the possibility to use the modulated potential as an effective tunable parameter for a
 system of interacting particles. 

In the present work the ground state configuration of a system of magnetic particles confined by
 a parabolic Q1D channel on a periodic substrate potential is studied. Magnetic particles are
 aligned with an external tilted magnetic field and interacting through an anisotropic
 dipole-dipole potential\cite{055_froltsov}, which depends on the orientation of the external
 field.  We investigate how the orientation of the field allows to tune the ground state
 configuration of the system, modulating the competitive interaction between each pair of
 particles.  The effect of the modulated potential is analyzed through its strength  and the
 degree of commensurability of the system imposed by the periodicity of the substrate potential
 and the average inter-particle distance, i.e. the density.

 The present paper is organized as follows. In Sec. II the model system and the numerical methods
 are described. In Sec. III we present the results for the different ground state transitions of
 the commensurate system, and the tuning between the different phases is realized by changing the
 orientation of the magnetic field and the external modulated potential.  In sect.~IV the same
 analysis is presented for a non-commensurate system, and the effects due to the commensurability
 factor are explained.  Our conclusions are given in Sec. V.

\section{Model System}

In this work, we consider a system of $N$ identical interacting magnetic particles, which are
 allowed to move in the $x-y$ plane.  The magnetic particles are confined by an external parabolic
 potential in the $y$ direction and a periodic substrate potential along the $x$ direction with
 period length $L$ and potential height $V$.

\begin{figure}[h!]
\begin{center}
\includegraphics[scale=0.50]{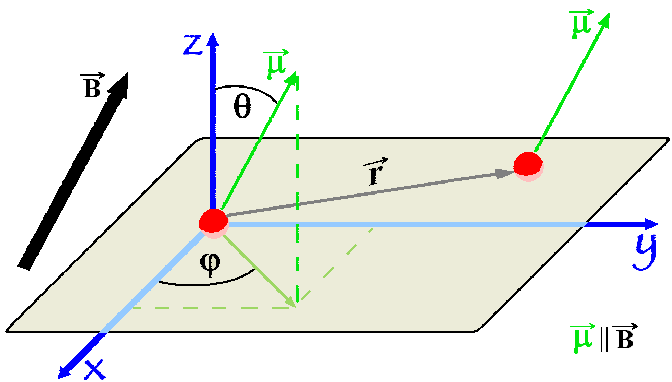}
\caption{\label{fig:schema} (Color online) Schematic representation of two dipolar particles (red
 spheres).  All particles are confined to the $x-y$ plane and the magnetic moment
 $\boldsymbol{\mu}$ of each one is aligned to the external magnetic field $\mathbf{B}$.}
\end{center}
\end{figure}

The system is subjected to an external and spatially homogeneous magnetic field $\mathbf{B}$,
 which induces a magnetic moment $\boldsymbol{\mu}$ to all particles, aligning them to
 $\mathbf{B}$.  The magnetic field is tilted by an angle $\theta$ with respect to the $z$-axis,
 which is perpendicular to the plane of motion, and it forms an angle $\varphi$ with respect to
 the $x$-axis of the plane.  This is schematically represented in Fig.~\ref{fig:schema}.
 
At short distances the interaction between the colloidal particles is a hard-core repulsion which
 defines the characteristic length (the diameter of each particle - $\sigma$) and the energy scale ($\varepsilon$) of the system.
 The total energy of the system is given by
\begin{eqnarray} \label{Energy_0}
  H & = & 
	  \sum_{i=1}^{N} \left[ \frac{1}{2}m\omega_{0}^{2}y_{i}^{2}
			       - V\cos\left(\frac{2\pi x_{i}}{L}\right) \right] \nonumber \\ & &
	+ \sum_{i=1}^{N} \sum_{j>i}^{N} \left[ 4\varepsilon\left( \frac{\sigma}{|\mathbf r_{ij}|} \right)^{12}
	  + V_{dip}(\mathbf r_{ij},\boldsymbol{\mu}) \right],
\end{eqnarray}
where $\mathbf{r}_{ij}=x_{ij}\mathbf{\hat{e}_{x}}+y_{ij}\mathbf{\hat{e}_{y}}$ represents the
 relative position between the $i$-th and the $j$-th particle and
 $V_{dip}(\mathbf r_{ij},\boldsymbol{\mu})$ is the dipole-dipole
 interaction\cite{055_froltsov, 084_lowen,  056_volosniev} given by
\begin{equation}\label{Vdip_0}
V_{dip}(\mathbf{r}_{ij},\boldsymbol{\mu}) = \frac{1}{r_{ij}^5}
	\left[ (\mu r_{ij})^2 - 3(\boldsymbol{\mu}\cdot\mathbf{r}_{ij})^2 \right],
\end{equation}
where we assume that the magnetic moment of each particle $\boldsymbol{\mu}$ is aligned to
 $\mathbf{B}$.  Equation~(\ref{Vdip_0}) can be rewritten as follows
\begin{eqnarray} \label{Vdip}
  V_{dip}(&\mathbf r_{ij}&,\theta, \varphi)
         =  \frac{\mu^{2}(1-3\sin^2{\theta}\sin^2(\varphi))}{|\mathbf r_{ij}|^3} \nonumber\\
         & - & \frac{3\mu^{2}\sin^2{\theta}}{|\mathbf r_{ij}|^5}x_{ij}
		      \left(\cos(2\varphi)x_{ij} - \sin(2\varphi)y_{ij}\right).
\end{eqnarray}

\begin{figure*}[htpb!]
\begin{center}
\includegraphics[trim={0.5cm 0.7cm 0 0}, scale=0.49]{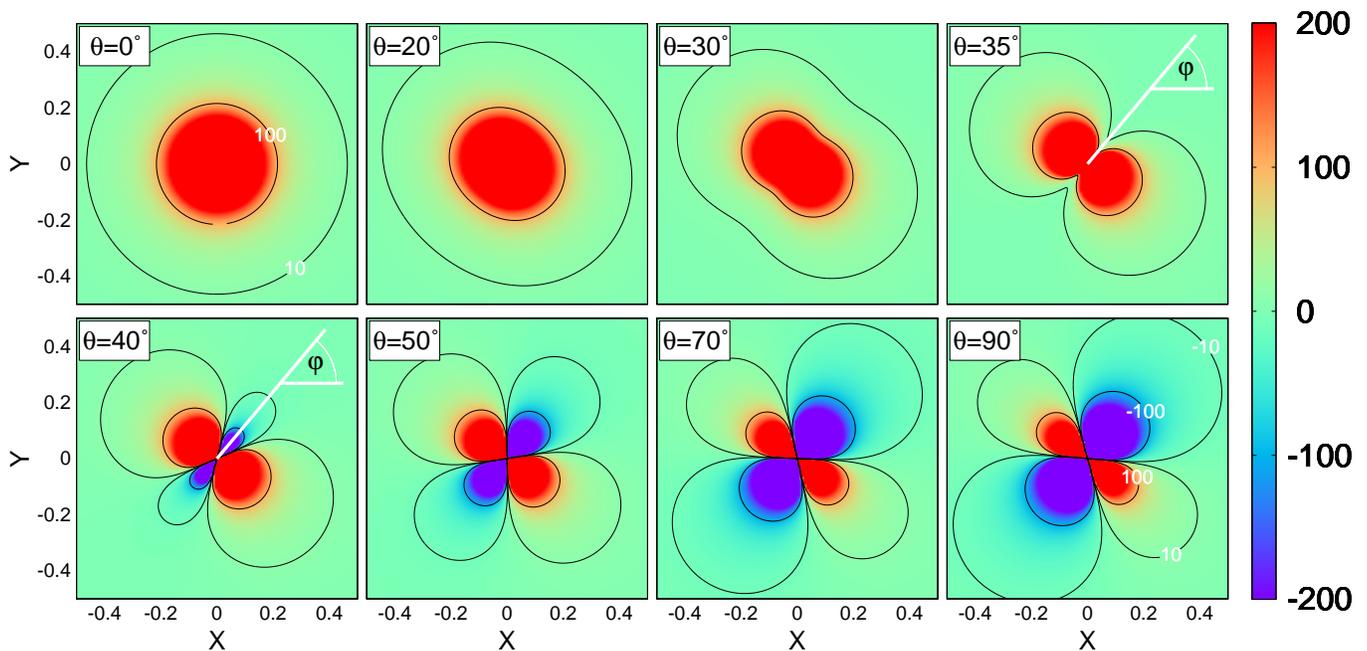}
\caption{\label{fig:dipole_interaction} (Color online) The dipole-dipole interaction as a function
 of the relative ($x$,$y$)-position between two particles, is plotted for different values of the
 angle $\theta$ between the magnetic field and the $z$-axis.  The value of $\theta$ is specified
 in the figures, while the strength of the interaction is indicated by the color bar. The contour
 lines are plotted for different values of the interaction energy, as labeled in the first and
 last figures.  The direction of $\mathbf{B}$ in the $x-y$ plane is indicated by the angle
 $\varphi$ in the top right figure.}
\end{center}
\end{figure*}

A contour plot of the pairwise interaction of Eq.~(\ref{Vdip}), is shown in
 Fig.~\ref{fig:dipole_interaction} as a function of the relative $x$ and $y$ distances between
 two particles for different values of $\theta$, as specified in the figures, with fixed
 $\varphi$.
The anisotropy of the dipole-dipole interaction (see Fig.~\ref{fig:dipole_interaction}), plays an
 important role in the determination of the many-particle ground-state configuration, as it will
 be discussed later, breaking the widely studied chain-like structure found for systems with
 isotropic pairwise interaction\cite{003_piacente, 041_galvan, 054_galvan}.  The anisotropy of the
 interaction is governed by the angles $\theta$ and $\varphi$, nevertheless, as shown in this
 work, the strength of the periodic potential can overcome the effect of the
 anisotropy, determining the many-particle configuration.

In dimensionless units, the total energy of Eq.~(\ref{Energy_0}) becomes
\begin{eqnarray} \label{Energy}
  H & = & 
	    \sum_{i=1}^{N} \Bigg[ \upsilon^{2}y_{i}^{2} 
				  - V_{0}\cos\left(\frac{2\pi x_{i}}{L}\right) \nonumber \\
    & + &   \sum_{j>i}^{N}
	    \bigg\{ 
	         \frac{4}{|\mathbf r_{ij}|^{12}}
		 + \frac{\mu^{2}(1-3\sin^2{\theta}\sin^2(\varphi))}{|\mathbf r_{ij}|^3}\nonumber\\
    & &	         \>\> - \frac{3\mu^{2}\sin^2{\theta}}{|\mathbf r_{ij}|^5}x_{ij}
		    \big(\cos(2\varphi)x_{ij} - \sin(2\varphi)y_{ij} \big)
	    \bigg\} \Bigg],
\end{eqnarray}
where the energy and the distances are given in units of $\varepsilon$ and $\sigma$, respectively.
 The dimensionless confinement frequency is defined by
 $\upsilon^{2}=m\omega_{0}^{2}/2\varepsilon\sigma^2$ and the dimensionless strength of the substrate
 potential by $V_{0}=V/\varepsilon$, while the dimensionless magnetic moment of each particle can
 be redefined by $\boldsymbol{\mu}/\varepsilon\sigma^{3} \Rightarrow \boldsymbol{\mu}$.

We introduce the dimensionless linear density $\eta$ and the periodicity number $n$, defined
 respectively as the number of particles and the number of periods of the periodic potential per
 unit length along the $x$ direction.  The presence of the periodic potential leads to the
 commensurability factor\cite{057_herrera}
\begin{equation} \label{comm_factor}
 p = \frac{N}{n} = \eta L,
\end{equation}
expressing the commensurability between the period $L$ and the average distance between the
particles.

In order to characterize the system, we define three different states of commensurability.
 $i$) $S_{I}$: when $p \in \mathbb{Z}$, which is a commensurate state; $ii$) $S_{II}$: when
 $p=k/m : k \in \mathbb{Z}$, $m \in \mathbb{N}$, $k \neq m$, it means that the system is not
 completely commensurate, this state could be subdivided into $S_{II}^{-}$ for $p<1$ and
 $S_{II}^{+}$ for $p>1$; $iii$) $S_{III}$: when $p \in \{\mathbb{R}-\mathbb{Q}\}$ and it implies that
 the system is incommensurate.  In this work we will limit our analysis to systems belonging to the first two
 categories.
 
We investigate the ground state configuration of the system, using Monte Carlo simulations (MC)
 optimized with the Newton method, which has been previously used to analyze the structural
 transitions in Q1D systems with purely repulsive
 interaction\cite{003_piacente, 006_piacente, 041_galvan, 054_galvan}. For all numerical results
 of the present work we consider $\upsilon=1$ and $\mu=1$.

We will study the dependence of the ground state configurations on the orientation of the magnetic
 field. For that, we fix the linear density of the system to the value $\eta=0.8$.
 Previously\cite{003_piacente, 006_piacente, 041_galvan} it was shown for $\mathbf{B}$
 perpendicular to the $x-y$ plane, the dipoles are no longer arranged in a single-chain
 configuration at the center of the parabolic channel for $\eta=0.8$, but split in a two-chain
 configuration.  The effect of the magnetic field strength will not be studied, since we are considering a
 system with superparamagnetic dipole particles, (i.e. the magnetic moment of all particles is
 aligned along $\mathbf{B}$).

\section{Commensurate System ($p=1$)} \label{p01o01}

\begin{figure}[h!]
\begin{center}
\includegraphics[trim={0.5cm 0 0 0}, scale=0.75]{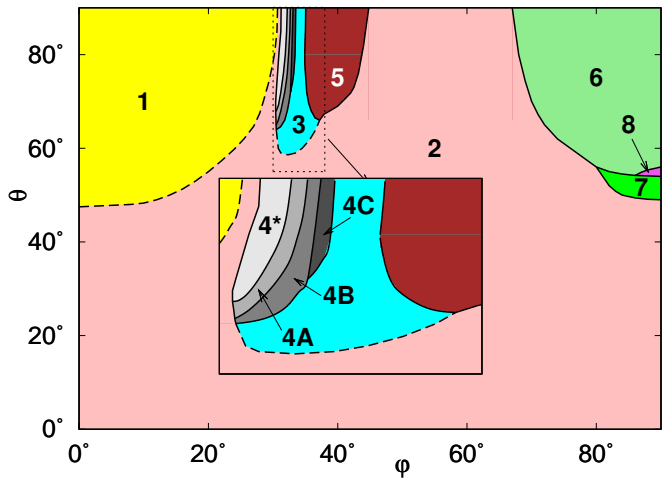}
\caption{\label{fig:p01o01_Vo_003} (Color online) Phase diagram of the ground state configuration,
 for a system of dipole particles in the regime $S_{I}$ with $p=1$ and $V_{0}=0.03$.  The
 solid (dashed) lines represent first (second) order transitions. Numbered phases are shown in
 Fig.~\ref{fig:p01o01_configs}.}
\end{center}
\end{figure}

For a typical commensurate state ($p=1$), the anisotropy of the system is strongly determined by
 the direction of $\mathbf{B}$.  In order to understand this effect, Fig.~\ref{fig:p01o01_Vo_003}
 shows the phase diagram of the system as a function of the planar ($\varphi$) and azimuthal
 ($\theta$) angles of the direction of $\mathbf{B}$, when the strength of the periodic potential
 is $V_{0}=0.03$.  In this diagram the solid and dashed curves represent first and second order
 phase transitions, respectively.

\begin{figure*}[htpb!]
\begin{center}
\includegraphics[scale=0.37]{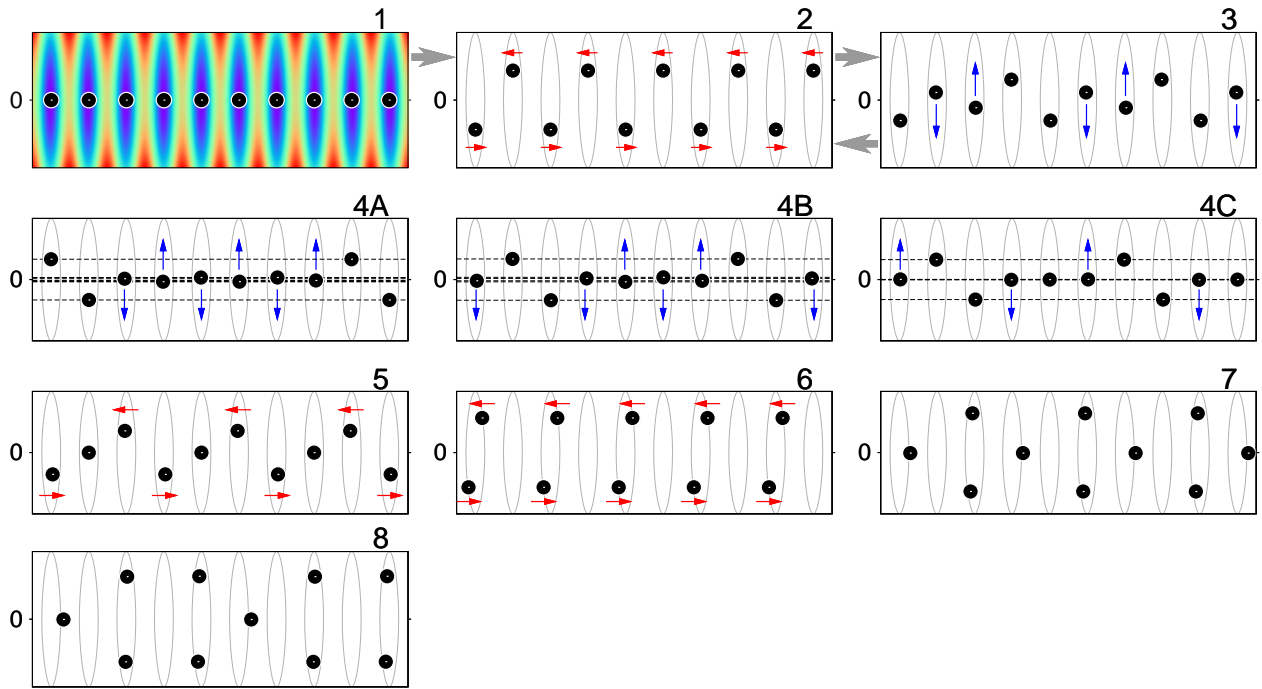}
\caption{\label{fig:p01o01_configs} (Color online) Different ground state configurations found
 for the system with $p=1$. The small red (blue) arrows show the movement of the particles in each
 phase caused by a small increment of $\varphi$ ($V_0$), while the gray ellipses represent the
 cells formed by the confining and periodic potentials (which is plotted as a color contour plot
 in $1$).}
\end{center}
\end{figure*}

In Fig.~\ref{fig:p01o01_configs} we present the ground state configurations as they are numbered
 in the phase diagram (see Fig.~\ref{fig:p01o01_Vo_003}).  The small red and blue arrows indicate the
 displacement of the particles when increasing $\varphi$ and $V_0$, respectively, while
 the gray arrows between two configurations indicate that the transition between them occurs
 continuously.
 
From Fig.~\ref{fig:p01o01_configs} one can see that, until phase $5$, all ground state
 configurations follow the following rule: one particle per one cell (we call cells the wells
 formed by the total confinement potential  which are marked in Fig.~\ref{fig:p01o01_configs} by
 gray ellipses).  However, with the exception of phase~$1$, when increasing $\varphi$ most of the
 particles move away from the center of the cell due to the anisotropic inter-particle
 interaction, following the directions shown by the red arrows.

\subsection{Transition from isotropy to anisotropy interaction}

From Eq.~(\ref{Vdip}) we notice that for $\theta=0^\circ$ the interaction between particles is
 purely repulsive and is given by $V_{dip}(r)=\mu^2/r^3$.  As was found in
 Refs.~[\onlinecite{003_piacente},\onlinecite{041_galvan},\onlinecite{006_piacente}], the ground
 state of the system is given by a zigzag configuration, which is represented by phase $2$.
 Notice also from Fig.~\ref{fig:p01o01_Vo_003}, that for $\theta=0^\circ$ the ground state is a
 zigzag configuration irrespective of the value of $\varphi$, evidencing the isotropy of the
 inter-particle interaction.
\begin{figure}[h!]
\begin{center}
\includegraphics[trim={0.5cm 0 0 0}, scale=0.75]{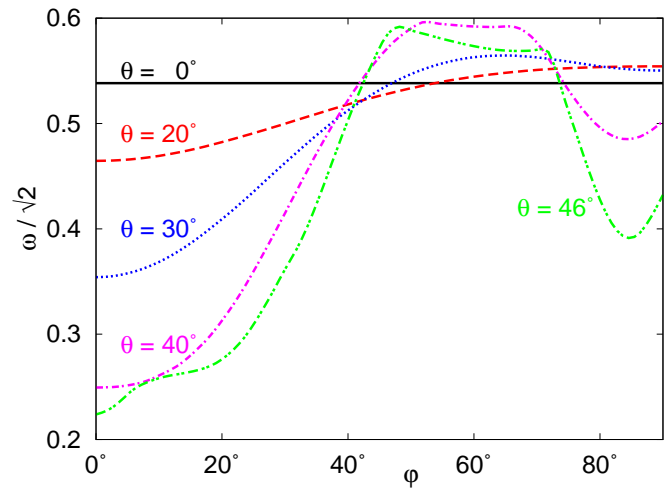}
\caption{\label{fig:p01o01_zigzag_freqs} (Color online) Lowest eigenfrequency of the two-chain
 configuration (phase 2) as a function of the planar angle ($\varphi$) for different values of
 $\theta$. The other parameters are $p=1$ and $V_{0}=0.03$.}
\end{center}
\end{figure}

The anisotropy of the interaction arises slowly by increasing $\theta$, resulting in a lowering of
 the lowest eigenfrequency of the system for small $\varphi$ and it slowly increases with
 $\varphi$ as seen in  Fig.~\ref{fig:p01o01_zigzag_freqs}.  Taking into account that the stability
 of a configuration is directly related to the value of its lowest eigenfrequency, we notice, that
 as an effect of the anisotropy of the pairwise interaction, the stability of the two-chain
 configuration (phase $2$ in Fig.~\ref{fig:p01o01_configs}) increases for intermediate values of
 $\varphi$ ($45^\circ \lesssim \varphi \lesssim 70^\circ$) by increasing $\theta$, while the
 opposite behaviour was found for $\theta \lesssim 30^\circ$.

\begin{figure}[h!]
\begin{center}
\includegraphics[trim={0.4cm 0.5cm 0 1.0cm}, scale=0.74]{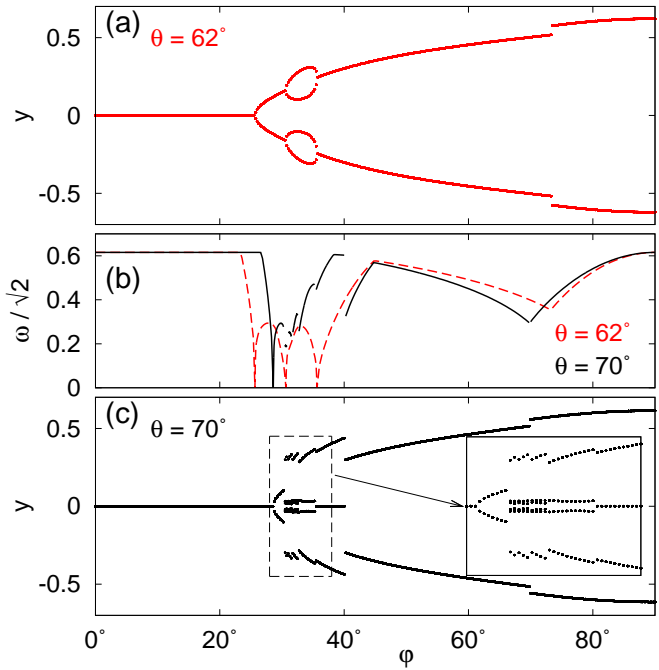}
\caption{\label{fig:p01o01_freqs} (Color online) (a), (c) The $y$-position of the particles in the
 ground state for a system with $p=1$ and $V_0=0.03$ as a function of $\varphi$, for two different
 values of $\theta$ indicated in each figure. (b) Lowest eigenfrequency of the system. Structural
 transitions occur when $\omega$ exhibits a jump (first order) or when it becomes zero (second
 order).}
\end{center}
\end{figure}

From Fig.~\ref{fig:p01o01_Vo_003} one can see that for $\theta \gtrsim 47^\circ$ the anisotropy of
 the pairwise interaction introduces new phases as ground state, now depending on $\varphi$. The
 very rich variety of phases produces in most of the cases first order transitions between them,
 as is evidenced in Fig.~\ref{fig:p01o01_freqs}, through the discontinuous jumps in the lowest
 eigenfrequency of the system.
 
The transition between the single- and the two-chain (phase~$2$) configuration (zigzag transition)
 appears in the region where the dipole-dipole interaction is dominantly attractive.  For
 $\theta \gtrsim 47^\circ$, the single-chain configuration (phase~$1$) is the ground state for
 small values of $\varphi$ and this region increases with increasing $\theta$. The continuous
 transition between these two phases in the $y$-direction, is shown in
 Fig.~\ref{fig:p01o01_freqs}(a), while Fig.~\ref{fig:p01o01_freqs}(b) shows that the lowest
 eigenfrequency goes to zero which is the evidence of a second order phase
 transition\cite{003_piacente}.

\subsection{Transition between $2-4-2$ chains: Re-entrant behavior}

A second order transition in the ground state configuration occurs between phases~$2$~and~$4$
 (Fig.~\ref{fig:p01o01_configs}).  Such a transition reveals an interesting effect which is put in
 evidence in Fig.~\ref{fig:p01o01_freqs}(a).  The transition between $2-4-2$ chains is found by
 increasing $\varphi$ within the interval ($59^\circ~\lesssim~\theta~\lesssim~63^\circ$)  in the
 $\theta-\varphi$ phase diagram (Fig.~\ref{fig:p01o01_Vo_003}).  This is demonstrated in
 Fig.~\ref{fig:p01o01_freqs}(a) for $\theta=62^\circ$. A similar re-entrant
 process has been found in dipolar fluids\cite{082_tlusty} and recently also for Q1D systems of
 patchy particles\cite{079_russo, 064_roldan} and ferrogels\cite{083_annunziata}.

This re-entrant process occurs after the zigzag transition takes place.  In
 Fig.~\ref{fig:p01o01_freqs}(b) we show that, by increasing $\varphi$, the two-chain configuration
 (phase~$2$) becomes less stable, since its lowest frequency approaches zero at the point where
 the four-chain configuration (phase~$3$) arises from a continuous lattice deformation of
 phase~$2$.  After that, the stability of the four-chain configuration increases until
 $\varphi \simeq 33^\circ$, where the lowest eigenfrequency  starts to decrease,
 returning to the two-chain configuration through a continuous deformation of the lattice.  During
 this process the lowest eigenfrequency of phase~$3$ decreases to zero, after that phase~$2$
 appears again through a second order transition (Fig.~\ref{fig:p01o01_freqs}(b)).

\begin{figure}[h!]
\begin{center}
\includegraphics[trim={0.35cm 0 0 0.8cm}, scale=0.44]{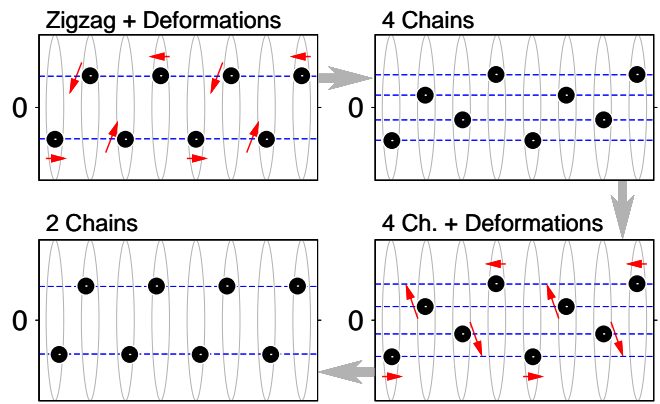}
\caption{\label{fig:p01o01_reentrant} (Color online) Mechanism of the re-entrant process between
 $2-4-2$ chains in the system.  The horizontal dashed lines accentuate the chain-like
 configuration of each phase.}
\end{center}
\end{figure}

The mechanism of this re-entrant process, which occurs through a continuous deformation of the
 lattice, is outlined in Fig.~\ref{fig:p01o01_reentrant}.
Notice that this re-entrant process is qualitatively different from the one found earlier in
 Ref.~\onlinecite{003_piacente}, where these processes were found to be of first order.  This
 difference is due to the presence of the periodic potential.  Without the periodic potential the
 particles not only undergo a zigzag transition, but they also exhibit a discontinuous shift
 in the $x$-direction which now is prevented by the periodic potential.

\subsection{Soliton-like deformations of the single-chain phase} \label{seq_deform}

For larger values of $\theta$ ($\theta\gtrsim70^\circ$) the anisotropy of the dipole-dipole
 interaction determines most of the transitions found for the ground state configurations.  It
 forces to align the particles along $\mathbf{B}$ forming (short) row-like arrangements, which may
 even result in the localization of two particles in the same cell of the periodic potential, as
 shown in Fig.~\ref{fig:p01o01_Vo_003} for $V_0=0.03$ (e.g., see phases~2,~5~and~6).
Some of these transitions result in a sequential deformation of the single-chain configuration as
 shown in the highlighted region in Fig.~\ref{fig:p01o01_freqs}(c), where the $y$-position of
 the particles is plotted as a function of $\varphi$ for $\theta=70^\circ$. In this case the
 deformations appear between the two- and four-chain configurations, and the region between the
 phases is identified in the phase diagram (Fig.~\ref{fig:p01o01_Vo_003}) as phase $4$.
 
In Fig.~\ref{fig:p01o01_freqs}(c) one can observe that, by increasing $\varphi$, the ground state
 phase transitions in the highlighted region are a sequence of single-chain and two-chain
 configurations followed by a group of disorder-like phases and after that four- and three-chain
 configurations, where all transitions are of first order with the exception of the zigzag
 transition which has been previously analyzed.  The first order transitions are evidenced by the
 discontinuities in the lowest eigenfrequency of the system\cite{003_piacente, 054_galvan} and
 the jumps in the $y$-coordinate position.
\begin{figure}[h!]
\begin{center}
\includegraphics[trim={0.35cm 0 0 0}, scale=0.72]{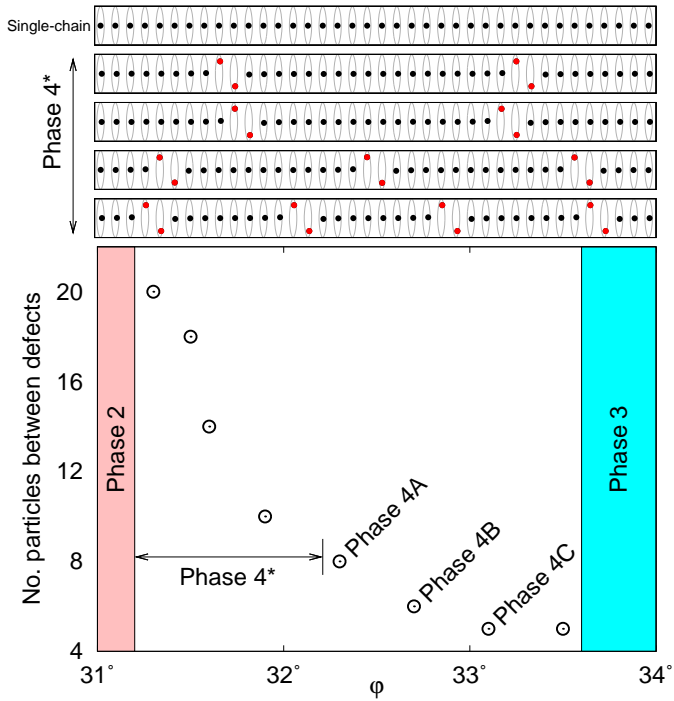}
\caption{\label{fig:p01o01_jumps} (Color online) The number of particles between defects (i.e.
 solitons) during the single-chain sequential deformation process, is plotted as a function of
 $\varphi$ for $\theta=90^\circ$ and $V_0=0.03$.  Configurations of the disordered phases are
 shown in the upper figures, where the positions of the defects are highlighted by red symbols for
 the particles, and the gray ellipses indicate the cells formed by the confinement in combination
 with the periodic potential.}
\end{center}
\end{figure}

The disorder-like phases aforementioned are a consequence of the high anisotropy of the
 interaction potential.  In order to understand this effect we consider the limiting case
 $\theta=90^\circ$, where the dipole-dipole interaction is dominantly
 attractive for $\varphi<30^\circ$.  In this limit the competition between the inter-particle
 interaction and the periodic confinement potential ($V_0=0.03$) produces a deformation of the
 single-chain configuration immediately after the two-chain configuration appears.  The
 single-chain configuration is suddenly broken by a local deformation of the lattice, after that,
 the period of the position where such a deformation occurs decreases with $\varphi$ until the
 four-chain configuration (phase $3$) is reached.

In the upper part of Fig.~\ref{fig:p01o01_jumps} we show the deformation process presenting
 some configurations that are typical for phase~$4^*$.  In those configurations the location of
 the defects are marked with red symbols for the particles.  They appear as frozen solitons whose
 density increases with $\varphi$.  From those configurations, the deformation process can be
 understood as a perturbation on the one-dimensional lattice, uniformly distributed on the axial
 direction, which allows that the system evolutes through first order transitions, into a Wigner
 crystalline structure.   In order to clarify this process, we plot the number of particles
 between defects (or solitons) as a function of $\varphi$ for the largest anisotropy of the
 interaction between particles (i.e. $\theta=90^\circ$) when $V_0=0.03$, in
 Fig.~\ref{fig:p01o01_jumps}.  This figure shows how the spatial frequency of the defects (or the
 density of solitons) increases by increasing $\varphi$, reducing the number of particles between
 defects until the ground state structure reaches the four-chain configuration (phase $3$).  This
 process has been found previously for an infinite system of particles interacting through a
 purely repulsive interaction, confined by a power-law potential $y^{\alpha}$ when $\alpha<2$, by
 increasing the linear density of the system\cite{041_galvan}.

\subsection{Effect of the periodic substrate potential}

\begin{figure}[h!]
\begin{center}
\includegraphics[trim={0.5cm 0.3cm 0 0}, scale=0.75]{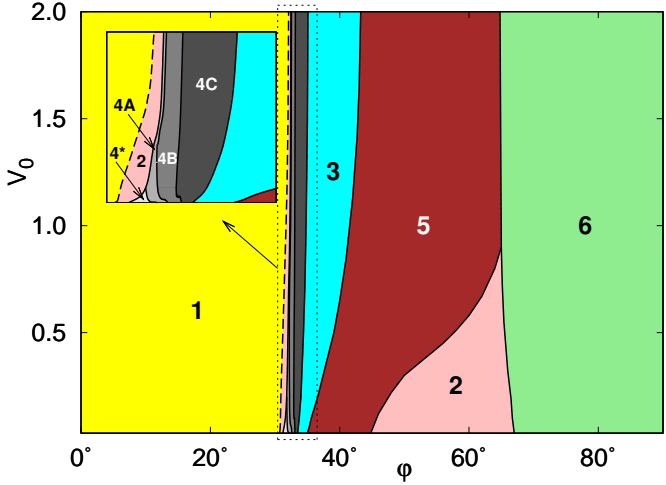}
\caption{\label{fig:p01o01_Th_900} (Color online) Phase diagram of the ground state configuration,
 for a system of dipole particles for the case $S_{I}$ with $p=1$ and $\theta=90^\circ$.
 Different phases are plotted in different colors as a function of the angle $\varphi$ and the
 strength of the substrate potential $V_0$.  The solid (dashed) lines represent first (second)
 order transitions between the phases. Numbered phases are shown in
 Fig.~\ref{fig:p01o01_configs}.  The dashed box area is enlarged in the inset.}
\end{center}
\end{figure}

In this section we pay attention to the effect of the substrate potential on the ground state
 configuration in the commensurate case with $p=1$ and $\theta=90^\circ$.  The ground state
 configurations are summarized in the $V_0-\varphi$ diagram which is presented in
 Fig.~\ref{fig:p01o01_Th_900}.  Notice that for $\theta=90^\circ$ the anisotropy of the
 dipole-dipole interaction is maximum (see Fig.~\ref{fig:dipole_interaction}).  For some intervals
 of $\varphi$ (e.g. $\varphi \lesssim 30^\circ$, $\varphi \gtrsim 68^\circ$) the ground state
 configuration is unaffected by $V_0$.  For $\varphi \lesssim 30^\circ$ the dipole-dipole
 interaction along the $x$-axis is dominantly attractive and the single-chain configuration is found
 as the ground state (Fig.~\ref{fig:p01o01_configs}).  On the other hand, for
 $\varphi \gtrsim 60^\circ$ the dipole-dipole interaction along the $x$-axis becomes dominantly
 repulsive and the two-chain configuration is now obtained as the ground state which is
 independent of $V_0$.  There is an intermediate interval
 $30^\circ \lesssim \varphi \lesssim 68^\circ$ in which it is possible to control the system
 configuration through the strength of the substrate $V_0$.

The process of the single-chain sequential deformation, as previously discussed, is attenuated by
 increasing $V_0$ reducing this process to a very narrow region (see inset in
 Fig.~\ref{fig:p01o01_Th_900}), where the deformation is reduced to the phases $4A-4C$.
 For $V_0>1$ the ground state configuration is less sensitive to a variation of $V_0$, as is
 visible in Fig.~\ref{fig:p01o01_Th_900}.  In all cases the effect of $V_0$ is to rearrange the
 configuration by moving the particles towards the center of the cell in the $y$-direction, as was
 previously studied in Ref.~[\onlinecite{018_carvalho}].
 
It is interesting to note that phases $7$ and $8$ do not appear in Fig.~\ref{fig:p01o01_Th_900},
 showing that these are characteristic configurations produced by the anisotropy of the
 interaction, which can be found in a small window of the angles $\theta$ and $\varphi$, for a
 weak strength of the periodic potential $V_0$, as shown in Fig.~\ref{fig:p01o01_Vo_003}.  The
 configuration of these phases (see Fig.~\ref{fig:p01o01_configs}) is evidence that the position
 of the particles are mainly determined by the interaction potential.

\section{Non-completely Commensurate System ($p=1/2$)} \label{p01o02}

\begin{figure}[h!]
\begin{center}
\includegraphics[trim={0.2cm 1.0cm 0 0}, scale=0.70]{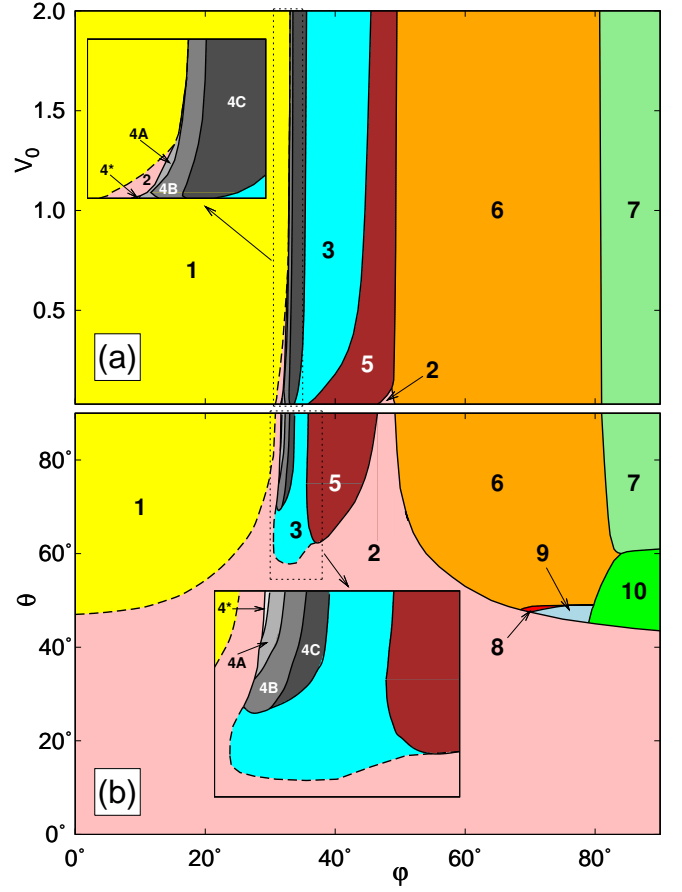}
\caption{\label{fig:p01o02_Vo_Th_Ph} (Color online) Phase diagram of the ground state
 configuration for a system of dipole particles in the regime $S_{II}$ with $p=1/2$.  Different
 phases are plotted as a function of (a) $\varphi$ and $V_0$ for $\theta=90^\circ$, and (b)
 $\varphi$ and $\theta$ for $V_{0}=0.03$.  The solid (dashed) lines represent first (second) order
 transitions between phases. Numbered phases are shown in Fig.~\ref{fig:p01o02_configs}.  The
 dashed box areas are elongated in the insets.}
\end{center}
\end{figure}

\begin{figure*}[htpb!]
\begin{center}
\includegraphics[scale=0.37]{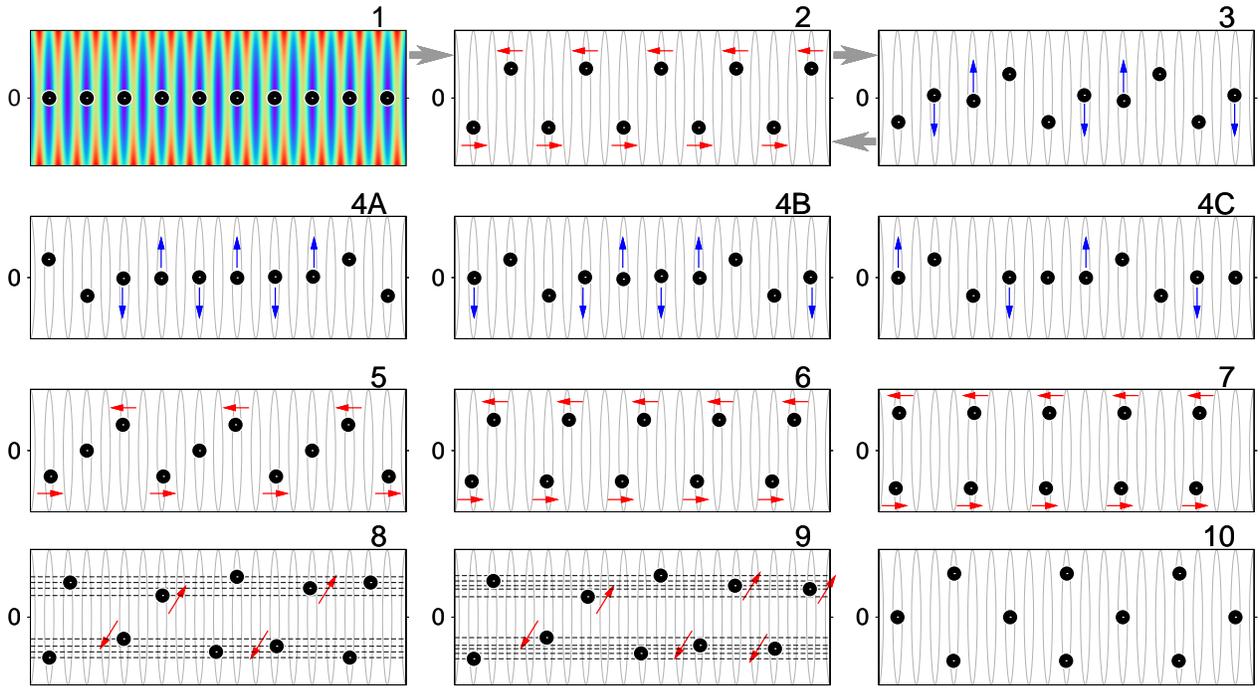}
\caption{\label{fig:p01o02_configs} (Color online) Different ground state configurations found
 for the system with $p=1/2$. The small red (blue) arrows show the movement of the particles in
 each phase caused by a small increment of $\varphi$ ($V_0$), while the gray ellipses represent
 the cells formed by the confining and periodic potentials (which is plotted as a color contour
 plot in $1$).}
\end{center}
\end{figure*}

We further analyze the influence of the commensurability factor on the system by taking $p=1/2$
 bearing in mind that $p$ can be controlled by $L$ or the linear density $\eta$
 (see~Eq.~(\ref{comm_factor})).  The ground state configurations are summarized in the
 $V_0-\varphi$ ($\theta-\varphi$) phase diagram for $\theta=90^\circ$ ($V_0=0.03$) which is
 presented in Fig.~\ref{fig:p01o02_Vo_Th_Ph}(a)~(\ref{fig:p01o02_Vo_Th_Ph}(b)).  The
 configurations of the numbered phases in Fig.~\ref{fig:p01o02_Vo_Th_Ph} are shown in
 Fig.~\ref{fig:p01o02_configs}, where the gray ellipses indicate the cells formed by the
 confinement and the periodic potential.

From Fig.~\ref{fig:p01o02_Vo_Th_Ph}(a), one can see that the effect of the periodic potential
 increases with increasing $V_0$, as was studied in previous section, and it results in a
 rearrangement of the configuration, moving the particles towards the center of the cell.  The
 single-chain sequential deformation process described by phases $4^*$ and $4A-4C$ which were
 analyzed in Sec. \ref{seq_deform}, is still present for $p=1/2$.  However, due to the existence
 of an empty cell between two nearest particles in these phases (see
 Fig.~\ref{fig:p01o02_configs}), the movement in the $x$-direction is highly restricted, as a
 consequence the single-chain deformation process will be reduced rapidly to the phases $4B$ and
 $4C$ by increasing $V_0$, as is shown by the inset in Fig.~\ref{fig:p01o02_Vo_Th_Ph}(a).  In the
 same way, the zigzag transition disappears around $V_0=0.5$, allowing the system to transit from
 phase~$1$ directly to phase~$4$ for higher values of $V_0$. 
 
On the other hand, Fig.~\ref{fig:p01o02_Vo_Th_Ph}(b) shows that due to the symmetry of the
 periodic potential wells, the effect of the interaction anisotropy on the system for small values
 of $\theta$, is similar to the previously discussed case of $p=1$ allowing to make the two-chain
 configuration (phase $2$ in Fig.~\ref{fig:p01o02_configs}) as the ground state.  Additionally, by
 increasing $\theta$ the stability of phase $2$ decreases allowing the system to reach other
 configurations as ground state. After that, several phases are found, but only until phase $5$
 the ground state configurations correspond to one particle per two cells, showing the large
 influence of $p$ on the system inside the region $\varphi<50^\circ$. 

As a consequence of the new symmetry imposed by the commensurability factor some new
 configurations, beyond those found for the case $p=1$, are found for the case $p=1/2$.
 As was discussed at the end of previous section, these configurations (phases~$6$,~$8$~and~$9$)
 are generated as a consequence of the anisotropy of the interaction potential for intermediate
 values of $\theta$.
\begin{figure}[htpb!]
\begin{center}
\includegraphics[trim={0.4cm 0.7cm 0 0}, scale=0.74]{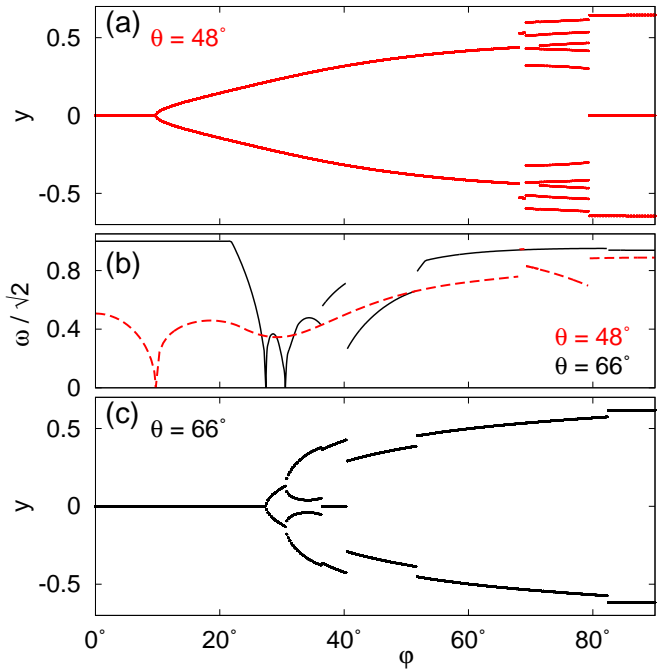}
\caption{\label{fig:p01o02_freqs} (Color online) (a), (c) The $y$-position of the particles in  the
 ground state for a system with $p=1/2$ and $V_0=0.03$ as a function of $\varphi$, for two
 different
 values of $\theta$ indicated in each figure. (b) Lowest eigenfrequency of the system.}
\end{center}
\end{figure}
 
Figs.~\ref{fig:p01o02_freqs}(a,b) show an interesting transition in the region
 $60^\circ \lesssim \varphi \lesssim 80^\circ$ for $\theta=48^\circ$ and $V_0=0.03$. In this
 region the ground state changes between phases $2-6-8-9-10$, where the first two correspond
 to different two-chain configurations, and the last one is a typical Q1D  Wigner crystal
 with a three-chain configuration.  Nevertheless, transition between phase $8$ (eight chains) and
 phase $9$ (ten chains) occurs with a small discontinuity in its lowest eigenfrequency as shown in
 Fig.~\ref{fig:p01o02_freqs}(b) around $\varphi=71^\circ$.  Such a transition is even more clear
 in Fig.~\ref{fig:p01o02_freqs}(a) where the $y$-position of the particles are plotted as a
 function of $\varphi$.  In Fig.~\ref{fig:p01o02_configs}, the configuration of phases~$8$~and~$9$
 are shown where the dashed lines indicate the position of the different chains.  Additionally,
 the red arrows show the displacement of the particles when $\varphi$ increases.

The two-chain configuration for $p=1/2$ is found in three different phases ($2$,~$6$~and~$7$ in
 Fig.~\ref{fig:p01o02_configs}),
 which shows that the zigzag symmetry is broken with increasing $\varphi$.
 However, the transition between these phases is discontinuous, even for small values of $V_0$,
 as shown in Fig.~\ref{fig:p01o02_freqs}(b) for $\theta=66^\circ$ and $V_0=0.03$, where the
 discontinuities in the lowest eigenfrequency of the system indicate first order transitions
 between these phases. These discontinuities are also evident in the $y$-position of the particles as
 a function of $\varphi$ in Fig.~\ref{fig:p01o02_freqs}(c).

\section{Conclusions}
We studied a Q1D system of magnetic particles confined by a parabolic channel modulated by a
 periodic substrate potential, where the magnetic moment of all particles is aligned by an
 external tilted magnetic field. The linear density in the direction of the modulated potential
 was fixed to $\eta=0.8$ and the ground state configuration at zero temperature was analyzed as
 a function of the magnetic field orientation, strength of the periodic potential and
 commensurability factor.  A plethora of different particle configurations are found as ground
 state, which are arranged in a different number of chains, and we even found a remarkable
 soliton-like configuration where the density of solitons could be varied with the tilt angle.

The anisotropy of the pairwise interaction between particles, determined by the magnetic tilt
 angles $\theta$ and $\varphi$ is largely responsible for the crystalline configuration of the
 system, but its effect decreases by increasing $V_0$.  For small values of $V_0$, the angle
 $\theta$ controls the degree of anisotropy of the system and with the in-plane angle $\varphi$
 it is possible to tune the ground state configuration.  On the other hand, for the limiting case
 of maximum pairwise anisotropy ($\theta=90^\circ$), the control of the ground state configuration
 depends weakly on $\varphi$ for large values of $V_0$.
 
The commensurability factor not only changes the stability region of the phases, as shown in the
 phase diagrams, but it also produces the emergence of characteristic ground state configurations
 due to its symmetry.  However, in order to get these characteristic phases for each
 commensurability factor, it is necessary to tune $\theta$ and $\varphi$ to appropriate values.

From these results we can conclude that varying $\theta$ and $\varphi$ allows us to tune the
 ground state of the system, while the effect of $V_0$ is to rearrange the configuration by moving
 the particles to the center of the cells in the $y$ direction.  The commensurability factor,
 controlled by $L$ or the density, acts as a complementary tunable parameter which gives the
 freedom to build a desired configuration.  This could be very useful as input parameter for
 experiments with magnetic particles, where the sinusoidal confinement can be realized by the
 application of external potentials through spatially varying light field, which have been used to
 induce structural changes in colloidal systems\cite{077_hanes, 063_drocco, 080_bechinger}, while
 the configurations can be obtained by video microscopy for different external magnetic fields as
 was demonstrated in Ref.~\onlinecite{084_lowen}.

\acknowledgments

This work was supported by the Flemish Science Foundation (FWO-Vl), the Methusalem programme of
 the Flemish government, CNPq, CAPES, FUNCAP (Pronex grant), the collaborative program
 CNPq~-~FWO-Vl and the Brazilian program Science Without Borders CsF.  Computational resources
 were provided by HPC infrastructure of University of Antwerp (CalcUA) a division of the Flemish
 Supercomputer Center (VSC).


%

\end{document}